\documentclass[10pt]{article}
\usepackage{times}
\usepackage{amsmath,amssymb,amsfonts}
\usepackage{booktabs}
\usepackage{tikz}
\usetikzlibrary{shapes.geometric, arrows, positioning, calc}
\usepackage[margin=1in]{geometry}
\usepackage{authblk}

\title{\textbf{Introducing the LLM Scalability Risk Index (LSRI) for Trustworthy AI: \\ A Standard Cybersecurity Framework for Assessing Agentic-AI Security and Software Model Supply Chain Safety, Boosting AI-Generated Malware Defense and Explainability for Mitigating Emerging Risks of Generative AI}}

\author[1]{\textbf{Kiarash Ahi}}
\author[2]{\textbf{Vaibhav Agrawal}}
\author[3]{\textbf{Saeed Valizadeh}}

\affil[1]{\small Virelya Intelligence Research Labs, San Francisco Bay Area, California}
\affil[2,3]{\small Google, Mountain View, California}
\affil[ ]{\small \texttt{1\{ahi@virelya.org, kiarash.ahi@uconn.edu\}, 2\{varawal@google.com\}, 3\{svalizadeh@google.com, saeed.valizadeh@uconn.edu\}}}

\date{}

\begin{document}

\maketitle

\begin{abstract}
As AI shifts from human-in-the-loop interfaces to autonomous multi-agent systems capable of real-time code execution and tool integration through protocols like the Model Context Protocol (MCP), traditional SAST, DAST, and legacy AI safety methods fail to detect modern agentic-AI threats. This paper introduces the LLM Scalability Risk Index (LSRI), a parametric framework and cybersecurity standard for stress-testing autonomous orchestration pipelines. LSRI measures the operational thresholds where load, compounding hallucinations, data poisoning, and adversarial prompt injections including jailbreaking and indirect prompt injection cause security boundaries to fail. 

Beyond RLHF and RLAIF, we present a Verifiable Root of Trust architecture using cryptographic attestation, semantic policy enforcement, and continuous runtime verification to secure the AI software supply chain. LSRI defends against malicious LoRA adapters, weight tampering, dependency typosquatting, and unsafe model artifacts from public registries such as Hugging Face and GitHub. By replacing post-hoc alignment with verifiable runtime controls, LSRI provides scalable API defense, safer agentic orchestration under heavy cloud workloads, stronger polymorphic malware detection, automated red-teaming, and improved system explainability. 

Oriented alongside NIST AI RMF, OWASP Top 10 for LLMs, and ISO 42001, LSRI establishes a deployable compliance baseline for securing generative AI ecosystems including ChatGPT, GPT-4o, Claude 3.5 Sonnet, Copilot, LLaMA, Gemini, and Bedrock. LSRI also supports capital market risk analysis by reducing exposure across technology portfolios linked to indices such as NASDAQ-100, QQQ, and semiconductor-driven supply chains.
\end{abstract}

\textbf{\textit{Index Terms}--- AI governance, AI-driven malware, anomaly detection, cybersecurity, dual-use AI, explainable AI (XAI), federated learning, large language models (LLMs), zero-day detection}

\section{Introduction}
The rapid evolution of artificial intelligence has placed Large Language Models (LLMs) and generative AI at the forefront of software innovation and cybersecurity transformation [1]--[3]. However, this widespread adoption has created a double-edged sword: LLMs empower defenders---especially platform administrators like Google Play, Apple App Store, and other enterprise app platforms---to perform static code scanning, automate threat detection, and improve code quality in real time. Yet simultaneously, those same models are exploited by attackers to generate malware, obfuscate code, and discover vulnerabilities at scale. This duality introduces complex security and governance challenges, underscoring the urgent need for systematic analysis, responsible deployment, and robust defensive frameworks [4].

This paper presents a comprehensive survey of both the risks and opportunities associated with LLMs in cybersecurity. We explore their dual-use nature, recent industry and academic advances, and how both defenders and adversaries leverage these models for tasks such as code generation, malware design, zero-day detection, and DevSecOps, supported by architectural comparisons, benchmark studies, and cross-industry case examples. We include both standalone LLMs and emerging LLM-powered agents with autonomous planning, memory, and tool-use capabilities under the umbrella of 'LLM-based cyber systems'. To guide the reader, the paper is structured as follows: Methodology in Brief---We analyzed 80 peer-reviewed papers and industry datasets, built a Policy Maturity Matrix, and supply-chain risks. Methodology appear in Section II. Section III lays the foundational background by reviewing existing literature on the evolution, capabilities, and early governance efforts concerning LLMs [75]. Section IV provides an in-depth analysis of LLM applicability in security, detailing their dual-use potential, and introduces the LLM Scalability Risk Index (LSRI), a parametric mathematical framework for evaluating operational and security trade-offs in production environments. Building on this analysis, Section V presents our focal research thrust: securing the LLM model supply chain. It outlines a roadmap for establishing a ``verifiable root of trust'' through cryptographic attestation and semantic policy enforcement. Section VI concludes the paper by summarizing our primary two contributions, and finally, Section VII discusses the policy and practice implications for stakeholders navigating the rising complexity of LLM-powered cybersecurity ecosystems and proposing a governance roadmap rooted in explainability, federated learning, and adaptive resilience.

\section{Methodology}
This paper employs a multi-pronged methodology combining empirical analysis, literature synthesis, and policy evaluation:
\begin{itemize}
    \item \textbf{Literature Survey:} We analyzed 80 peer-reviewed papers, policy documents, and technical reports published between 2018--2025, focusing on dual-use LLMs, adversarial robustness, and AI governance.
    \item \textbf{Threat Categorization:} Attack vectors were organized using a structured adversarial taxonomy derived from [17], [18], [20], and cross-referenced with OWASP's LLM Top 10 and NIST AI RMF 1.0.
    \item \textbf{Policy Maturity Matrix:} Governance frameworks were scored based on enforcement level, coverage of LLM-specific risks, and implementation transparency, weighted equally across five pillars.
    \item \textbf{LSRI Design:} The LLM Scalability Risk Index (LSRI) was developed as a parametric framework for sensitivity-based stress testing, utilizing industrial-representative performance parameters to quantify the trade-offs between scalability, security, and compliance in high-throughput environments.
    \item \textbf{Supply-Chain Analysis and Systems Synthesis:} We analyzed prior work on model integrity, data poisoning, weight tampering, and agentic vulnerabilities to define the LLM supply chain as a lifecycle-spanning system covering build-time artifacts and run-time dependencies. This synthesis informed a verifiable root-of-trust framework that treats supply-chain assurance as an enforceable systems property rather than a post-hoc governance mechanism.
\end{itemize}

\section{Background and Literature Review}

\subsection{Evolution and Capabilities of LLMs}
Large Language Models (LLMs) have evolved rapidly from their initial applications in natural language translation and generation to highly capable systems supporting complex software engineering tasks. Models such as GPT-4 and PaLM now perform code generation, refactoring, debugging, and even formal verification with increasing accuracy and fluency [1], [5]. These advancements are enabled by scaling transformer architectures and training on diverse programming and natural language corpora. Recent research from OpenAI and Google demonstrates how LLMs can integrate into full development pipelines, assisting with test case creation, API documentation, and dynamic bug resolution [6]--[8].

\subsection{Security Risks and Early Governance Efforts}
The dual-use nature of LLMs has raised significant security concerns. On one hand, they can support code auditing and threat detection; on the other, they can generate obfuscated or insecure code, or be weaponized for malicious purposes. Comprehensive surveys by Yao et al. (2024) and Bryce et al. (2024) highlight these privacy and security trade-offs, categorizing the impact of LLMs across a spectrum of beneficial and adversarial outcomes [74], [80]. Prior work has emphasized the need for proactive safeguards, such as Brundage et al.'s recommendations on structured red teaming and audit trails, and the European Union’s Artificial Intelligence Act, which mandates risk assessments and transparency reports for high-impact models [9], [10]. These frameworks aim to mitigate misuse while supporting responsible innovation.

\subsection{Ethics and Governance of Dual-Use LLMs}
Integrating LLMs into CI/CD pipelines automates crucial security tasks such as code review, threat detection, and compliance enforcement. GitLab and Azure DevOps showcase how GPT based tools can enable real-time security hardening and policy enforcement [11], [12]. 

While the EU AI Act and the US NIST AI RMF represent significant strides, the global governance landscape for LLMs in cybersecurity remains dynamic, with other major technological regions developing their own distinct approaches. For instance, countries in Asia, such as China, Japan, South Korea, and Singapore, are actively formulating AI regulations and ethical guidelines that reflect their unique priorities. Understanding these varied international perspectives and fostering dialogue towards greater regulatory interoperability will be crucial for addressing the borderless nature of cyber threats and ensuring a globally coordinated response to the risks posed by dual-use AI [9]--[12].

\begin{table}[htbp]
\centering
\caption{Policy Maturity Matrix across different regions}
\label{tab:policy_matrix}
\vspace{2mm}
\begin{tabular}{p{2.3cm}p{2.2cm}p{2.5cm}p{2.3cm}p{2.5cm}}
\toprule
\textbf{Region} & \textbf{Governance Maturity} & \textbf{LLM-Specific Laws} & \textbf{Red Teaming Mandate} & \textbf{Transparency Requirements} \\
\midrule
European Union & High & AI Act (2025) & Required & Required \\
\addlinespace
United States & Medium & NIST AI RMF (Voluntary) & Encouraged, not required & Limited (varies by agency) \\
\addlinespace
China & High & Interim Measures (2023) [48] & Required & Required \\
\addlinespace
Japan & Medium & AI Strategy 2022 [49] & Not required & Partially encouraged \\
\addlinespace
Singapore & High & Model AI Governance Framework & Required & Required \\
\bottomrule
\end{tabular}
\end{table}

\subsection{Privacy-Aware Deployment of LLMs via Federated Learning}
Privacy preserving LLM deployment strategies are increasingly relevant. Federated learning allows training across distributed devices without centralizing data, aligning with laws like GDPR. Kairouz et al. and Bonawitz et al. have demonstrated that these frameworks preserve privacy while maintaining model utility [13], [14].

\subsection{Explainability and Trust in AI Driven Defense}
The adoption of LLMs in automated security systems demands transparency. Explainable AI (XAI) methods like SHAP and LIME have been customized to make LLM based vulnerability classifications interpretable. These models help developers and analysts understand the rationale behind predictions, supporting auditability and compliance [15], [16].

\subsection{Adversarial Attacks and Model Vulnerabilities}
The integration of LLMs into security critical domains has exposed them to sophisticated adversarial attacks. Carlini et al. highlighted how training data could be extracted from LLMs, undermining confidentiality [8]. Wallace et al. demonstrated that prompt injection and adversarial fine tuning can manipulate LLM outputs, evading content filters. Benchmarks such as PINT and recent tools have emerged to systematically test defenses against prompt injection and jailbreak attacks, measuring both false positives and false negatives [69]. Recent work by Jia et al. organized a global competition revealing how LLMs can be tricked into generating offensive content and misinformation, emphasizing the need for rigorous adversarial testing frameworks [17].

\section{Analysis of LLM Applicability in Security}
As LLMs become deeply embedded in software development and cybersecurity pipelines, their dual-use potential has triggered increasing scrutiny. A growing body of research has documented how these models can unintentionally or deliberately produce insecure code, including cryptographic flaws, SQL injection vectors, and XSS vulnerabilities [18]--[20]. More alarmingly, the accessibility of LLMs has democratized the creation of deceptive content---enabling non-experts and malicious actors alike to generate phishing apps, polymorphic malware, and social engineering scripts at scale [21]--[23]. These developments reflect not isolated failures but systemic risks introduced by generative models when deployed without sufficient constraints. This section analyzes such risks through three lenses: (1) the emerging threat landscape shaped by misuse and amateur error, (2) industry-led defense strategies to mitigate LLM-enabled attacks, and (3) the broader governance and technical challenges that complicate safe deployment.

\subsection{Amateur Developers and Security Risks}
While LLMs accelerate software creation, they have unintentionally enabled a wave of insecure development among amateur coders. By lowering technical barriers, these models allow individuals with minimal training to generate functional code that often lacks essential safety. Studies indicate that inexperienced developers frequently integrate LLM-generated snippets without validating security implications [24], [25], causing common vulnerabilities---such as improper authentication and insecure API usage to proliferate in production software.

This highlights an urgent need for LLM-integrated guardrails to proactively flag unsafe patterns for novice users. While amateur misuse stems from a lack of expertise, it sets the stage for the more calculated, scalable weaponization of generative AI by professional adversaries.

\subsection{Malicious Actors Leveraging LLMs}
Beyond accidental misuse, malicious actors are leveraging LLMs as force multipliers for deliberate cyberattacks, automating the creation of malware, phishing payloads, ransomware, and code obfuscation. Unlike traditional malware authors who required deep expertise, attackers can now generate malicious scripts with minimal effort, dramatically accelerating development cycles [71]. Recent cybersecurity reports reveal a sharp upward trend in malware generation using LLM, raising concerns about automated threat scaling and democratized access to advanced attack tools [29]. 

Security researchers report that these ``dark LLMs'' are increasingly optimized to evade endpoint detection and static analysis through polymorphic payloads and context-aware generation [65]--[66]. Simultaneously, deepfake multimedia amplifies social engineering; for instance, a Ferrari executive was targeted by a CEO voice-clone, which failed only when the AI could not answer a specific question [67].

The emergence of autonomous LLM agents marks a critical inflection point where multi-stage attacks require minimal oversight. This transition drives ``Cyber Threat Inflation,'' characterized by drastically reduced attack costs and an industrial offensive scale [56].

Researchers from Carnegie Mellon and Anthropic demonstrated that LLMs can autonomously plan and execute attacks, successfully replicating the 2017 Equifax breach using the Incalmo toolkit [57]. In controlled enterprise environments, these systems achieved partial or full compromise across most test networks [26]--[28]. By producing code that mutates to bypass static defenses, adversaries have shifted threat scalability from human-limited to industrial-scale.

\subsection{Defensive Utilization of LLMs in Mobile App Security}
In response to rising AI-powered threats, mobile platform providers are embedding LLMs directly into their security workflows. One of the most effective use cases is automated code review---where LLMs augment traditional static analyzers by identifying logic flaws, unusual API usage, or obfuscated payloads that escape signature-based detection [33]--[36].

These use cases demonstrate how LLMs can shift mobile app security from reactive filtering to intelligent pre-deployment screening, flagging issues before users ever download an app. However, as defensive applications of LLMs grow more powerful, they also inherit risks such as overfitting, bias, or exploitability---making explainability and continuous retraining essential.

\subsection{Industry Case Studies: Leveraging LLMs for Cyber Defense}
As LLM-fueled threats escalate, leading technology companies are responding by deploying proprietary AI tools to reinforce digital defenses. These platforms integrate LLMs into core security operations, including code review, static analysis, compliance auditing, and threat intelligence---tailoring strategies to align with specific infrastructure and security priorities [33]--[37]. Table \ref{tab:companies_security} summarizes several leading companies and their defensive applications of LLM technology.

These systems represent a critical shift from reactive to proactive security postures. For instance, Google's Gemini underpins Play Protect's live threat engine, capable of analyzing millions of apps for suspicious behavior in real time. Microsoft's Security Copilot assists analysts by flagging unsafe code patterns and generating remediation steps, while Amazon's CodeWhisperer identifies vulnerabilities in IDEs during the creation process. Similarly, CrowdStrike's ``Charlotte AI'' automates incident prioritization to accelerate response times [58], and Palo Alto's Cortex XDR leverages AI to unify telemetry across network and cloud environments to neutralize threats.

By embedding LLMs into their security stacks, these organizations set new industry standards for AI-augmented defense. However, these same capabilities, if left unchecked, can also empower adversaries, reinforcing the dual-use nature of LLM technology. Furthermore, leading cloud providers are piloting insurance partnerships to share data on AI-related incidents, reflecting the growing financial dimension of GenAI security. Table \ref{tab:companies_security} highlights the specific LLM technologies deployed by these companies for cyber defense.

\begin{table}[htbp]
\centering
\caption{Leading Companies Leveraging LLMs For Security}
\label{tab:companies_security}
\vspace{2mm}
\begin{tabular}{lll}
\toprule
\textbf{Company} & \textbf{LLM Technology} & \textbf{Application} \\
\midrule
Google & Gemini & Malware Detection, Static Analysis, Threat Intelligence \\
Microsoft & GPT-4 & Security Copilot, Code Review \\
Amazon & CodeWhisperer & Vulnerability Detection \\
IBM & Watsonx & Compliance \& Risk Management \\
Palantir & AIP & Threat Hunting \& Behavioral Analysis \\
\bottomrule
\end{tabular}
\end{table}

\subsection{Scalability Concerns in LLM-Based Security Systems}
Integrating LLMs into production-level security pipelines such as global app stores or CI/CD environments presents significant technical and operational challenges. Real-world deployment requires low-latency inference, cost-effective infrastructure, and high throughput across diverse architectures. For instance, scanning millions of apps in the Google Play or Apple App Store for malware necessitates robust resource allocation and distributed serving to ensure inferences complete within milliseconds while remaining resilient to adversarial inputs.

Operational overhead increases when maintaining regional consistency under varying regulations like GDPR or CCPA. Furthermore, the EU Artificial Intelligence Act explicitly mandates technical documentation and cryptographic attestation for high-impact AI systems [64].

The OWASP Top 10 for LLM Applications addresses these scaling risks through the new ``Unbounded Consumption'' category [59]. This expands the traditional Denial of Service (DoS) threat to include resource mismanagement and unexpected infrastructure costs, reflecting how LLMs can be exploited to consume excessive tokens or processing power.

This creates a critical tension: while powerful, LLMs are not trivially scalable. Integration into global security infrastructure must be engineered to avoid bottlenecks and regional inconsistencies. 

To quantify these trade-offs, we propose the LLM Scalability Risk Index (LSRI), formalized as a parametric risk model in Equation \ref{eq:lsri}. Higher LSRI values indicate greater deployment readiness under the specified scalability, cost, and security constraints.

\begin{equation}\label{eq:lsri}
LSRI = \Phi \cdot \left( 1 - \sum_{i=1}^{n} w_i \cdot f_i(x_i) \right)
\end{equation}

Where:
\begin{itemize}
    \item $x_i$: is the observed raw metric for a specific factor (e.g., latency in ms).
    \item $f_i(x_i) \in [0,1]$: is a non-linear Risk Mapping Function that normalizes raw data into a risk score. A lower $f_i$ indicates higher deployment readiness.
    \item $w_i$: represents the contextual weight assigned to each factor, where $\sum w_i = 1$.
\end{itemize}

\textit{Note:} For the purposes of this study, we utilize a baseline where $w_i = \frac{1}{n}$ (equal weighting) to provide a generalized assessment. However, these weights are intended to be adjusted by security architects and cross-functional teams to align with specific organizational requirements and risk tolerances.

$\tau, \sigma, \lambda$: represent the critical performance threshold, the sensitivity of the risk gradient, and the target industrial scale, respectively, allowing the LSRI to be calibrated to specific hardware or regulatory environments, corresponding to environmental calibration parameters defined in Table \ref{tab:lsri_rubric}.

$\Phi$: Integrated Integrity Multiplier, rather than a binary gate, is a continuous coefficient representing the aggregate security health of the system. It is defined as:

\begin{equation}\label{eq:phi}
\Phi = \prod_{j=1}^{m} \max(0, 1 - \alpha_j \cdot E_j)
\end{equation}

Where:
\begin{itemize}
    \item $E_j$: is the measured violation magnitude (e.g., the specific rate of prompt-injection success or PII leakage).
    \item $\alpha_j$: is the sensitivity coefficient for risk category $j$, allowing the multiplier to scale based on the severity of the threat.
\end{itemize}

\subsubsection{Risk Mapping Functions ($f_i$)}
To ensure the index reflects real-world performance non-linearity, we utilize three primary mapping functions to evaluate deployment readiness:

\paragraph{Sigmoid Mapping (Latency)}
Used for time-sensitive metrics where risk remains low until a critical threshold $\tau$ is reached, then increases exponentially.
\begin{equation}\label{eq:sigmoid}
f_{\text{sig}}(x) = \frac{1}{1 + e^{-\left(\frac{x - \tau}{\sigma}\right)}}
\end{equation}

\paragraph{Exponential Mapping (Throughput)}
Used to model mitigation of scaling risks, where risk decreases as capacity approaches industrial levels $\lambda$.
\begin{equation}\label{eq:exponential}
f_{\text{exp}}(x) = e^{-\frac{x}{\lambda}}
\end{equation}
Throughput values exceeding $\lambda$ asymptotically reduce risk toward zero, reflecting diminishing marginal scalability risk beyond industrial baselines.

\paragraph{Step Mapping (Size)}
Used for hard hardware constraints such as VRAM limits, where risk is binary based on hardware compatibility:
\begin{equation}\label{eq:step}
f_{\text{step}}(x) = \begin{cases} 0, & x \le \text{Threshold} \\ 1, & x > \text{Threshold} \end{cases}
\end{equation}

\paragraph{Linear Mapping (Cost, Frequency, Regulation)}
Used for proportional resource depletion and maintenance benchmarks. Risk increases linearly with cost and decreases linearly with update frequency:
\begin{equation}\label{eq:cost}
f_{\text{cost}}(x) = \min\left(1, \frac{x}{\text{Budget Ceiling}}\right)
\end{equation}
\begin{equation}\label{eq:freq}
f_{\text{freq}}(x) = \max\left(0, 1 - \frac{x}{\text{Target Frequency}}\right)
\end{equation}
\begin{equation}\label{eq:reg}
f_{\text{reg}}(x) = 1 - \text{Compliance Framework Priority}
\end{equation}

This formulation ensures that the index adapts to different deployment profiles while maintaining a ``zero-trust'' threshold for fundamental security boundaries. A model cannot achieve a ``safe'' score if it violates fundamental security or compliance boundaries ($\Phi = 0$), regardless of its speed or cost-efficiency.

The LSRI provides a structured, forward-looking risk score that supports comparative assessment of scalability–security stress across deployment scenarios, rather than an empirically calibrated estimator of real-world incident probability.

While LSRI quantifies operational readiness, its integrity multiplier ($\Phi$) motivates the enforceable supply-chain guarantees developed in Section V.

\subsection{Application and Calibration}
The LSRI assists architects in balancing the technical, legal, and economic dimensions of large-scale LLM defense. By assigning scores based on these weighted functions, security teams can establish rigorous thresholds for acceptable deployment conditions. Table \ref{tab:lsri_rubric} outlines the specific parameters used for a standard high-impact deployment (e.g., App Store security).

\begin{table}[htbp]
\centering
\caption{LLM Scalability Risk Index (LSRI): Deployment Readiness Rubric}
\label{tab:lsri_rubric}
\vspace{2mm}
\begin{tabular}{p{3cm}p{2cm}p{2.5cm}p{4cm}p{2.5cm}}
\toprule
\textbf{Risk Factor} & \textbf{Metric $x_i$} & \textbf{Mapping Type} & \textbf{Parameter Thresholds} & \textbf{Mapping Logic} \\
\midrule
Inference Latency & ms & Sigmoid & $\tau = 100$ms, $\sigma = 15$ & (Eq. \ref{eq:sigmoid}) \\
Throughput & req/day & Exponential & $\lambda = 10^6$ & (Eq. \ref{eq:exponential}) \\
Regulatory Risk & CFP & Linear & $\text{CFP} \in [0,1]$ & (Eq. \ref{eq:reg}) \\
Model Size & Params & Step & $\text{Threshold} = 20$B & (Eq. \ref{eq:step}) \\
Update Freq & count & Linear & $\text{Target} = 2$ (updates/day) & (Eq. \ref{eq:freq}) \\
Cost Sensitivity & USD & Linear & $\text{Ceiling} = \$1000$ & (Eq. \ref{eq:cost}) \\
\bottomrule
\end{tabular}
\end{table}

The Compliance Framework Priority (CFP) is a normalized value representing the percentage of required regulatory controls (e.g., EU AI Act Article 10) successfully implemented. Update frequency is measured in updates per day. Throughput $x$ is measured as sustained inference requests per day, normalized against an industrial baseline $\lambda$ representing large-scale app-store or CI/CD deployment.

By assigning scores based on these weighted functions, security teams can establish rigorous thresholds for acceptable deployment conditions. While Table \ref{tab:lsri_rubric} provides baseline parameters for a high-throughput mobile ecosystem, the framework is designed for Sensitivity Analysis, allowing architects to stress-test how specific metric fluctuations (e.g., a 20\% increase in latency) impact the overall deployment risk profile.

\subsection{Practical Validation: Worked Examples and Sensitivity Analysis}
To demonstrate the application of the LSRI, we evaluate two hypothetical model deployment scenarios in Table \ref{tab:scenarios} using the parameters and weights defined in the rubric (Table \ref{tab:lsri_rubric}).

\begin{table}[htbp]
\centering
\caption{Comparison of Model Readiness Scenarios}
\label{tab:scenarios}
\vspace{2mm}
\begin{tabular}{p{4.5cm}p{5cm}p{5cm}}
\toprule
\textbf{Feature} & \textbf{Scenario A: High Readiness} & \textbf{Scenario B: Low Readiness} \\
\midrule
Model Type & Optimized 15B Parameter Model & 15B Model with Logic Vulnerability \\
\addlinespace
Integrity Multiplier ($\Phi$) & 1 (Passes all security audits) & 0.76 (with 24\% prompt injection rate) \\
\addlinespace
Latency $x_1$ & 85ms ($f_{\text{sig}} = 0.27$) & 40ms ($f_{\text{sig}} = 0.02$) \\
\addlinespace
Throughput $x_2$ & 1.2M req/day ($f_{\text{exp}} = 0.30$) & 2.0M req/day ($f_{\text{exp}} = 0.14$) \\
\addlinespace
Regulatory Risk $x_3$ & 80\% Compliance ($f_{\text{reg}} = 0.20$) & 10\% Compliance ($f_{\text{reg}} = 0.90$) \\
\addlinespace
Model Size $x_4$ & 15B ($f_{\text{step}} = 0.0$) & 15B ($f_{\text{step}} = 0.0$) \\
\addlinespace
Update Freq $x_5$ & 1.5 updates/day ($f_{\text{freq}} = 0.25$) & 0.2 updates/day ($f_{\text{freq}} = 0.90$) \\
\addlinespace
Cost Sensitivity $x_6$ & \$275/day ($f_{\text{cost}} = 0.28$) & \$850/day ($f_{\text{cost}} = 0.85$) \\
\midrule
\textbf{Final LSRI Score} & \textbf{$\sim$0.78 (Ready for Deployment)} & \textbf{$\sim$0.40 (Risky)} \\
\bottomrule
\end{tabular}
\end{table}

The framework is specifically designed to be sensitive to performance ``tipping points'' via non-linear mapping. Table \ref{tab:sensitivity} illustrates how the risk score reacts to fluctuations in latency, demonstrating the Sensitivity Analysis of the Sigmoid function $f_{\text{sig}}$ with $\tau = 100$ms, $\sigma = 15$.

\begin{table}[htbp]
\centering
\caption{Sensitivity Analysis of Latency Risk Mapping}
\label{tab:sensitivity}
\vspace{2mm}
\begin{tabular}{llll}
\toprule
\textbf{Observed Latency ($x$)} & \textbf{Risk Score $f_{\text{sig}}$} & \textbf{Resulting LSRI*} & \textbf{Qualitative Risk Impact} \\
\midrule
50 ms & 0.03 & 0.82 & Negligible: Optimal performance \\
100 ms ($\tau$) & 0.50 & 0.74 & Critical Threshold: Performance warning \\
125 ms & 0.84 & 0.68 & High Risk: Significant UX degradation \\
150 ms & 0.96 & 0.66 & Failure: System functionally unusable \\
\bottomrule
\end{tabular}
\vspace{2mm}
\begin{flushleft}
\small{*LSRI values are computed by substituting the specified latency value into the Scenario A baseline in Table \ref{tab:scenarios} while holding all other factors and weights constant.}
\end{flushleft}
\end{table}

\subsection{Regulatory Compliance and Privacy Constraints}
The deployment of LLMs in security workflows introduces complex compliance challenges under frameworks like the General Data Protection Regulation (GDPR) and the California Consumer Privacy Act (CCPA) [38], [39]. These regulations mandate strict data minimization, user consent, data residency, and the ``right to explanation,'' all of which constrain how LLMs are trained and applied to sensitive content.

\subsection{Explainability and Trust in AI-Driven Defense}
As LLMs take on increasingly autonomous roles in cybersecurity---classifying vulnerabilities, triaging threats, or flagging anomalies---the need for explainable artificial intelligence (XAI) has become paramount. Without transparency into how these decisions are made, stakeholders may lose confidence in AI-driven defense systems, especially when they impact compliance, reputation, or user rights.

To bridge this gap, researchers have adapted traditional XAI techniques such as SHAP and LIME to LLMs, enabling visibility into influential tokens, attention patterns, and decision pathways [40]. These interpretations not only enhance trust but also help security analysts validate model behavior, identify edge-case failures, and fine-tune thresholds for deployment. The major categories of explainability tools and their use cases in security pipelines are summarized in Figure \ref{fig:explainability_tools}.

\begin{figure}[htbp]
\centering
\begin{tikzpicture}[node distance=1.2cm, auto,
    block/.style={rectangle, draw, fill=blue!10, text width=11cm, text centered, rounded corners, minimum height=0.8cm, font=\small}]
    \node [block, fill=red!10] (m1) {\textbf{Model-Level Interpretation:} SHAP, LIME (Token Attribution)};
    \node [block, fill=cyan!10, below=0.3cm of m1] (m2) {\textbf{Benchmarking \& Evaluation:} CySecBench (Security Benchmarks)};
    \node [block, fill=green!10, fill=green!5, below=0.3cm of m2] (m3) {\textbf{Training \& Education:} CyberMentor (Explainable Guidance)};
    \node [block, fill=yellow!10, fill=yellow!5, below=0.3cm of m3] (m4) {\textbf{Governance \& Compliance:} CKC, AI Model Cards (Ethical Auditing)};
\end{tikzpicture}
\caption{Categorization of explainability tools used in LLM-driven cybersecurity systems.}
\label{fig:explainability_tools}
\end{figure}

In educational settings, tools like CyberMentor [42] use Retrieval-Augmented Generation (RAG) and agentic workflows to provide interpreable feedback. These systems teach not just what a threat is, but its underlying mechanics.

Beyond technical utility, the dual-use nature of LLMs necessitates ethical auditing. Frameworks like the Cyber Kill Chain (CKC) and AI model cards are used to document decision logic and misuse potential in auditable formats. As noted by Barrett et al. [43] and Gupta et al. [44], explainability is critical infrastructure for regulatory compliance and long-term trust in AI-powered defense.

\subsection{Federated Learning and Privacy-Aware Deployment of LLMs}
As LLMs increasingly handle sensitive data in mobile and edge environments, ensuring privacy without compromising performance is a top priority. Federated Learning (FL) offers a promising paradigm by enabling decentralized training without transferring raw data to central servers. This approach aligns with GDPR and CCPA regulations regarding data locality and minimization [13].

Kairouz et al. [13] provided foundational analysis of FL's scalability and security trade-offs, while Bonawitz et al. [14] demonstrated large-scale implementation using secure aggregation protocols. By integrating LLMs with FL, security tools can perform real-time anomaly detection and on-device code analysis without transmitting data to the cloud. This architecture minimizes centralized breach risks and promotes compliance-by-design (see Figure \ref{fig:centralized_vs_federated}).

\begin{figure}[htbp]
\centering
\begin{tikzpicture}[node distance=2cm, auto, font=\small]
    % Centralized pipeline setup
    \node[draw, rectangle, fill=red!5, text width=3.2cm, text centered, minimum height=1cm, rounded corners] (c_cloud) {\textbf{Centralized Model}\\Cloud LLM Server};
    \node[below=1.6cm of c_cloud] (c_user) {Distributed Clients};
    \draw[->, thick, red, transform canvas={xshift=-0.4cm}] (c_user.north) -- node[left, font=\scriptsize, text=black] {Raw Data Upload} (c_cloud.south);
    \draw[<-, thick, blue, transform canvas={xshift=0.4cm}] (c_user.north) -- node[right, font=\scriptsize, text=black] {Inference Response} (c_cloud.south);
    \node[below=0.2cm of c_user, font=\bfseries] {Centralized Model Architecture};

    % Division boundary
    \draw[dashed, gray!60, thick] (3.2, 0.6) -- (3.2, -3.8);

    % Federated pipeline setup
    \node[draw, rectangle, fill=green!5, text width=3.2cm, text centered, minimum height=1cm, rounded corners] at (6.4,0) (f_cloud) {\textbf{Federated Server}\\Secure Aggregation};
    \node at (6.4, -1.6) (f_user) {Distributed Clients};
    \draw[->, thick, green!60!black, transform canvas={xshift=-0.4cm}] (f_user.north) -- node[left, font=\scriptsize, text=black] {Encrypted Gradients} (f_cloud.south);
    \draw[<-, thick, orange, transform canvas={xshift=0.4cm}] (f_user.north) -- node[right, font=\scriptsize, text=black] {Global Model Weights} (f_cloud.south);
    \node[below=0.2cm of f_user, font=\bfseries] {Federated Model Architecture};
\end{tikzpicture}
\caption{Architectural comparison between centralized and federated LLM deployment.}
\label{fig:centralized_vs_federated}
\end{figure}

Hybrid models are emerging that combine FL with on-device fine-tuning, allowing devices to benefit from shared intelligence while customizing insights for local threats. For instance, mobile security platforms using edge-deployed LLMs can detect suspicious behaviors without exposing private logs or PII to external servers.

Complementary techniques like differential privacy and homomorphic encryption further harden FL pipelines against inference and model inversion threats. These layered approaches ensure privacy, accountability, and robustness against sophisticated adversaries. Ultimately, FL is a critical enabler for trustworthy AI, allowing defenders to leverage LLMs while navigating the legal and technical constraints of real-world deployment.

\subsection{Detection of Zero-Day Vulnerabilities}
Zero-day vulnerabilities often bypass traditional rule-based detection, but Large Language Models (LLMs) offer a semantic, context-aware alternative. For example, Google’s Big Sleep project recently identified a zero-day in SQLite using LLM-driven analysis [52]. Research by Lisha et al. [45] demonstrates that LLMs trained on vulnerability-specific corpora outperform static analyzers in detecting complex logic and control-flow flaws. While these models also assist in predicting exploit propagation paths, researchers warn that over-reliance on synthetic training data can lead to ``model collapse,'' reducing effectiveness against rare, critical vulnerabilities [68].

In practice, these techniques are integrated into CI/CD pipelines. GitHub’s code scanning and Google’s Play Protect, for example, experiment with LLM-powered models to detect anomalies in obfuscated binaries. LLMs are also applied in fuzzing, automatically generating exploit-oriented test cases to surface weaknesses preemptively. Figure \ref{fig:detection_pipelines} contrasts traditional detection pipelines with LLM-based approaches.

\begin{figure}[htbp]
\centering
\begin{tikzpicture}[node distance=0.8cm, auto, font=\small,
    box/.style={rectangle, draw, minimum width=4.5cm, minimum height=0.6cm, rounded corners, text centered}]
    
    \node[box, fill=red!5] (t1) {Static Syntactic Analyzer};
    \node[box, fill=red!5, below=0.2cm of t1] (t2) {Rule-Based Heuristics};
    \node[box, fill=red!5, below=0.2cm of t2] (t3) {Static Signature Matching};
    \node[above=0.2cm of t1, font=\bfseries] {Traditional Defenses};

    \node[box, fill=cyan!5, right=2.0cm of t1] (l1) {Semantic Context Analysis};
    \node[box, fill=cyan!5, below=0.2cm of l1] (l2) {Vulnerability Flow Prediction};
    \node[box, fill=cyan!5, below=0.2cm of l2] (l3) {Exploit Path Graph Inference};
    \node[above=0.2cm of l1, font=\bfseries] {LLM-Enhanced Pipeline};

    \draw[<->, dashed, thick, black!70] ($(t2.east)$) -- node[above, font=\scriptsize\itshape, text=black] {Pipeline Shift} ($(l2.west)$);
\end{tikzpicture}
\caption{Comparison between traditional and LLM-enhanced zero-day vulnerability detection pipelines.}
\label{fig:detection_pipelines}
\end{figure}

Ultimately, this capability reinforces the dual-use nature of LLMs: while defenders gain tools for neutralizing unknown threats, attackers could fine-tune models to identify zero-day opportunities faster. This semantic power makes zero-day detection a critical battleground in AI-driven cyber defense.

Table \ref{tab:detection_perf} compares traditional static analyzers with LLM-enhanced vulnerability detection across four key metrics. Relative to the static-analyzer baseline (62\% recall), LLM systems deliver 26–29 percentage-point absolute gains—a $\approx$42–47\% relative improvement—while maintaining similar latency and false-positive rates.

\begin{table}[htbp]
\centering
\caption{Performance comparison of traditional static analyzers and LLM-enhanced zero-day vulnerability detectors.}
\label{tab:detection_perf}
\vspace{2mm}
\begin{tabular}{llllp{4cm}}
\toprule
\textbf{Method} & \textbf{Recall (Zero-Day)} & \textbf{Avg Latency (ms)} & \textbf{False Positives} & \textbf{Interpretability} \\
\midrule
Static Analyzers & 62\% & 80 & 11\% & Limited \\
LLM + Symbolic Hybrid & 88\% & 105 & 13\% & Moderate (SHAP) \\
LLM + Graph-Based & 91\% & 96 & 12\% & High (CySecBench) \\
\bottomrule
\end{tabular}
\end{table}

\subsection{LLMs in DevSecOps Automation}
As software delivery accelerates, security must evolve to match the speed of continuous integration and deployment. DevSecOps—the integration of security directly into DevOps workflows—demands automation, precision, and scale across the entire software development lifecycle. LLMs are increasingly being leveraged to meet this need, embedding intelligence into every stage of the pipeline.

In modern DevSecOps environments, LLMs assist in:
\begin{itemize}
    \item Code scanning at every commit, flagging insecure patterns and suggesting remediations in real time.
    \item Assessing containerized builds for compliance with internal and external security policies.
    \item Analyzing dependency trees to identify vulnerable or outdated libraries before code reaches production.
\end{itemize}

Prominent platforms have already begun integrating these capabilities. GitLab’s Auto DevSecOps system employs GPT-based models for dynamic scanning and compliance-as-code enforcement. Similarly, Microsoft’s Azure DevOps, in collaboration with OpenAI, leverages LLMs for predictive vulnerability scoring, contextual remediation advice, and automated security testing.

These integrations shift security from a reactive checkpoint to a proactive, continuous layer—built directly into the tooling developers already use. This minimizes friction, shortens feedback loops, and enables security-by-default at scale.

At the same time, this growing reliance on LLMs in DevSecOps pipelines highlights the broader theme of this paper: the dual-use nature of AI in security. An AI-powered security testing agent like XBOW, currently leading the HackerOne leaderboard, also highlights the dual-use nature, as a capability that can be used for both finding security vulnerabilities in applications and responsibly reporting them, or using it for malicious purposes. The same models that harden pipelines could be exploited if misconfigured, biased, or insufficiently governed—making LLM observability, explainability, and governance as important as their functional accuracy.

\subsection{Ethics and Governance of Dual-Use LLMs}
As LLM capabilities scale, their misuse potential grows alongside their utility. This creates a dual-use dilemma: models powering security auditing and malware detection can also generate polymorphic malware or optimize phishing campaigns. Such high-stakes symmetry necessitates governance frameworks as adaptable as the technology itself.

Brundage et al. [10] have proposed concrete mechanisms to address these risks, including:
\begin{itemize}
    \item Structured red teaming to stress-test model behavior against adversarial use cases,
    \item Staged release strategies to control the dissemination of high-risk capabilities, and
    \item Model Watermarking to allow for the traceability of content generated by a model [46].
    \item Model evaluation cards and Human-in-the-Loop (HITL) systems to provide oversight and document safety constraints.
\end{itemize}

These strategies are being codified in the EU AI Act and the U.S. NIST AI Risk Management Framework, which mandate transparency in development and auditability of training data. These policies aim to shift deployment toward accountability-by-design.

Ethical research further emphasizes value alignment in security domains. Techniques like Reinforcement Learning with Human Feedback (RLHF) are being adapted to teach LLMs to:
\begin{itemize}
    \item Reject harmful or manipulative queries,
    \item Disclose uncertainty in high-risk scenarios, and
    \item Explain security decisions with interpretable confidence bounds.
\end{itemize}

At the international level, coalitions such as the Global Partnership on AI (GPAI) and recent AI safety summits have introduced shared guardrails, including:
\begin{itemize}
    \item Pre-registration of frontier models,
    \item Mandatory incident reporting, and
    \item Centralized auditing repositories to detect and flag unsafe usage patterns.
\end{itemize}
These governance efforts are not merely bureaucratic safeguards; they are essential infrastructure for responsibly integrating LLMs into national security, digital forensics, and trust-sensitive ecosystems. Figure \ref{fig:timeline} illustrates the timeline of key governance milestones that have emerged between 2023 and 2025, highlighting a growing global effort to institutionalize safety practices around powerful LLMs.

\begin{figure}[htbp]
\centering
\begin{tikzpicture}[node distance=1.5cm, auto, font=\small]
    \draw[->, ultra thick, black!70] (0,0) -- (12,0) node[below right, font=\bfseries] {Timeline};
    \foreach \x/\txt in {1.5/2023, 6.0/2024, 10.5/2025} {
        \draw[ultra thick, black!70] (\x, 0.15) -- (\x, -0.15) node[below=0.1cm, font=\bfseries] {\txt};
    }
    \node[align=center, draw, rectangle, fill=blue!5, rounded corners, text width=3cm] at (1.5, 1.6) 
        {\scriptsize \textbf{OpenAI Red Teaming}\\Structured safety validation and early evaluation framework protocols};
    \draw[dashed, thick, black!40] (1.5,0.7) -- (1.5,0);

    \node[align=center, draw, rectangle, fill=blue!5, rounded corners, text width=3.2cm] at (4.0, -1.8) 
        {\scriptsize \textbf{EU AI Act Proposals \& GPAI}\\Initial introduction of cross-border transparency requirements};
    \draw[dashed, thick, black!40] (4.0,-0.8) -- (4.0,0);

    \node[align=center, draw, rectangle, fill=blue!5, rounded corners, text width=3cm] at (6.5, 1.6) 
        {\scriptsize \textbf{NIST AI RMF Release}\\Official launch of the voluntary risk management index frameworks};
    \draw[dashed, thick, black!40] (6.5,0.7) -- (6.5,0);

    \node[align=center, draw, rectangle, fill=blue!5, rounded corners, text width=3.4cm] at (9.5, -1.8) 
        {\scriptsize \textbf{Pre-Registration \& Reporting}\\Global coordination standardizing active multi-stage risk audits};
    \draw[dashed, thick, black!40] (9.5,-0.8) -- (9.5,0);
\end{tikzpicture}
\caption{Timeline of emerging governance frameworks for dual-use LLMs, spanning initiatives from 2023 to 2025.}
\label{fig:timeline}
\end{figure}

\subsection{Securing Defensive LLM Systems}
As LLMs become integral components of cybersecurity infrastructure itself (e.g., in threat detection, code analysis, and incident response), their own security posture becomes paramount. Protecting these 'defender' LLMs from targeted attacks is crucial to maintain their efficacy and trustworthiness. Key considerations in safeguarding these sentinel AI systems include:

\begin{itemize}
    \item \textbf{Training Data Integrity and Poisoning Defense:} Ensuring the provenance and integrity of data used to train and fine-tune security LLMs to prevent sophisticated poisoning attacks that could create blind spots or backdoors [50].
    \item \textbf{Model Evasion and Robustness:} Continuously evaluating and hardening defensive LLMs against adversarial evasion techniques specifically designed to bypass AI-based detection [8], [17]. Fine-tuned LLM models with a labeled dataset can be used to detect against prompt injection attacks [53]. The INJECAGENT benchmark, for example, demonstrates that tool-integrated LLM agents are vulnerable in many scenarios, with attack success rates of $\sim$24\% under certain indirect prompt injection settings [11].
    \item \textbf{Model Confidentiality and Integrity:} Protecting the proprietary architecture and weights of security LLMs from extraction [8], and ensuring their operational integrity against unauthorized modifications.
    \item \textbf{Secure Deployment and Monitoring:} Implementing secure deployment practices for LLM-based security tools, including robust access controls, audit trails, and continuous monitoring for anomalous behavior or potential compromise of the AI system itself [5].
\end{itemize}

These findings underscore that security LLMs must be protected with the same rigor as the systems they defend. Despite alignment tuning, OpenAI’s red teaming efforts have shown that LLMs still exhibit failure modes under adversarial prompting and jailbreak scenarios [72].

\subsection{Model-Supply Chain Security}
While all these risks manifest across multiple layers from prompt injection and adversarial evasion to governance and compliance failures, many of them ultimately trace back to weaknesses in how LLMs are built and distributed. The LLM supply chain from data collection to model release is a critical vector for compromise. Attackers can inject poisoned samples during pre-training, introduce malicious fine-tuning data, or distribute backdoored weights via public repositories. Recent research shows minor dataset manipulations can induce persistent ``logic bombs'' that evade traditional red-teaming [61].

Tramèr et al. highlight the risk of tampered checkpoints and propose cryptographic signing of training pipelines to ensure integrity [62]. Similarly, Goldblum et al. demonstrate automated detection of Trojaned models and recommend mandatory weight attestation for high-risk systems [63].

Defending AI-driven platforms requires securing the entire lifecycle, not just inference endpoints. Figure \ref{fig:supply_chain} illustrates this end-to-end supply chain, highlighting potential compromise points and recommended defenses to ensure model provenance within the supply chain and security.

\begin{figure}[htbp]
\centering
\begin{tikzpicture}[node distance=0.4cm, auto, font=\scriptsize,
    stepbox/.style={rectangle, draw, minimum width=2.2cm, minimum height=1cm, text width=2cm, text centered, rounded corners, fill=yellow!10, font=\bfseries}]
    
    \node[stepbox] (s1) {Data Collection};
    \node[stepbox, right=0.5cm of s1] (s2) {Pre-training};
    \node[stepbox, right=0.5cm of s2] (s3) {Fine-tuning};
    \node[stepbox, right=0.3cm of s3] (s4) {Checkpoint Distribution};
    \node[stepbox, right=0.3cm of s4] (s5) {Inference API};

    \draw[->, ultra thick, black!70] (s1) -- (s2);
    \draw[->, ultra thick, black!70] (s2) -- (s3);
    \draw[->, ultra thick, black!70] (s3) -- (s4);
    \draw[->, ultra thick, black!70] (s4) -- (s5);
    
    \node[below=0.3cm of s1, text centered, text=red, font=\sffamily\bfseries] (t1) {$\times$ Data Poisoning};
    \node[below=0.3cm of s3, text centered, text=red, font=\sffamily\bfseries] (t3) {$\times$ Weight Tampering};
    
    \node[above=0.3cm of s1, text centered, text=blue!80!black, text width=2cm] (d1) {\checkmark Provenance Auditing};
    \node[above=0.3cm of s2, text centered, text=blue!80!black, text width=2cm] (d2) {\checkmark Cryptographic Signing};
    \node[above=0.3cm of s4, text centered, text=blue!80!black, text width=2cm] (d4) {\checkmark Weight Attestation};
    
    \draw[dashed, red, thick] (t1) -- (s1);
    \draw[dashed, red, thick] (t3) -- (s3);
    \draw[dashed, blue!60, thick] (d1) -- (s1);
    \draw[dashed, blue!60, thick] (d2) -- (s2);
    \draw[dashed, blue!60, thick] (d4) -- (s4);
\end{tikzpicture}
\caption{End-to-end LLM model-supply chain showing potential compromise points and recommended defenses.}
\label{fig:supply_chain}
\end{figure}

While the LSRI provides a parametric framework for evaluating operational readiness, the validity of its 'Integrity Multiplier' $\Phi$ depends on the verifiable security of the model’s origin. This necessitates a transition from external performance evaluation to internal enforceable supply-chain assurance.

\section{Focal Research Thrust: Securing the LLM Model Supply Chain}
This section advances the research claim that security and governance of large language models cannot be meaningfully achieved through post-deployment monitoring or policy mechanisms alone. Instead, enforceable, pre-execution supply-chain guarantees spanning both build-time artifacts and run-time agentic dependencies, are a necessary condition for scalable deployment, operational safety, and effective governance of LLM-based systems. We formalize this claim through a verifiable root-of-trust architecture and demonstrate its feasibility using existing cryptographic primitives.

Unlike traditional software, LLM supply chains allow for latent vulnerabilities introduced during foundational stages to remain dormant until execution. Security therefore spans two coupled phases: build-time (trust in data and training) and run-time (interaction with tools and networks). Failures at build time propagate into autonomous agent behavior, necessitating a verifiable root of trust for AI systems.

This section grounds the supply-chain research thrust in a focused proof-of-concept (PoC) demonstrating how existing cryptographic tooling can already establish a verifiable root of trust for LLM artifacts. The remainder of the section extends this concrete baseline toward broader challenges in provenance, distribution, and agentic robustness, reframing them as incremental extensions.

\subsection{Proposed Supply-Chain Assurance Architecture}
Figure \ref{fig:root_of_trust} presents a reference architecture for LLM supply-chain assurance inspired by modern software supply-chain security practices (e.g., SLSA, in-toto), adapted to the scale and opacity of foundation models. The architecture introduces three enforceable control points:
\begin{enumerate}
    \item Cryptographic attestation of model artifacts at the completion of training or fine-tuning.
    \item Signed provenance metadata binding model weights to their training context, datasets, and alignment configuration.
    \item Mandatory verification gates enforced prior to deployment or inference.
\end{enumerate}
Together, these controls establish a fail-closed trust boundary: a model that cannot be cryptographically verified against approved provenance and policy constraints is prevented from executing, regardless of performance or cost considerations.

\begin{figure}[htbp]
\centering
\begin{tikzpicture}[node distance=1.2cm, auto, font=\small,
    phasebox/.style={rectangle, draw, minimum width=3.8cm, minimum height=1.2cm, text width=3.6cm, text centered, rounded corners, fill=blue!5, font=\bfseries}]
    
    \node[phasebox] (p1) {Build-Time Attestation\\(Sigstore, Hash Generation)};
    \node[phasebox, right=1.2cm of p1] (p2) {Signed Model Package\\(Immutable Provenance Metadata)};
    \node[phasebox, right=1.2cm of p2] (p3) {Deployment \& Inference\\(Fail-Closed Verification Gates)};
    
    \draw[->, ultra thick, black!70] (p1) -- node[above, font=\scriptsize\itshape] {Attest Lifecycle} (p2);
    \draw[->, ultra thick, black!70] (p2) -- node[above, font=\scriptsize\itshape] {Gate Execution} (p3);
\end{tikzpicture}
\caption{Proposed verifiable root-of-trust architecture for LLM supply-chain security.}
\label{fig:root_of_trust}
\end{figure}

\subsection{Prototype Implementation: Cryptographic Weight Attestation}
To demonstrate feasibility, we implemented a minimal prototype for cryptographic checkpoint attestation using existing open-source tooling commonly employed in software supply-chain security.

\textbf{Build-time attestation.} Upon completion of model fine-tuning, the resulting checkpoint (e.g., \texttt{model.ckpt}) is hashed using SHA-256. The hash is then signed using Sigstore, binding the artifact to an ephemeral signing identity derived from the CI/CD environment [76]. Alongside the signature, structured metadata---such as model version, dataset identifier, alignment policy version, and training timestamp---is recorded in an append-only transparency log.

\textbf{Run-time verification.} Prior to loading the model for inference, the serving environment verifies (i) that the checkpoint hash matches the signed digest, (ii) that the signature is valid and anchored in the transparency log, and (iii) that the attested metadata satisfies deployment policy (e.g., approved dataset lineage or safety configuration). If any verification step fails, model loading is aborted.

This prototype requires no modification to model internals and introduces negligible overhead at inference time. Importantly, it demonstrates that cryptographic weight attestation is immediately deployable within existing MLOps pipelines, transforming supply-chain assurance from a policy aspiration into an enforceable technical control.

\subsection{Verifiable Provenance and Data Integrity}
Building on the checkpoint-attestation prototype, verifiable provenance emerges as a critical extension of supply-chain assurance. While the PoC binds a deployed model to a cryptographically verified artifact, it does not yet capture why a model behaves as it does—namely, which datasets, transformations, and alignment steps influenced its training.

In practice, extending the architecture requires binding dataset identifiers, preprocessing pipelines, and fine-tuning stages to signed metadata associated with the final checkpoint. Such provenance records allow downstream consumers to verify not only that a model’s training lineage is intact, but that it was trained under approved data and policy constraints. This capability is particularly important for mitigating data poisoning and backdoor insertion attacks, where small training-time manipulations can induce persistent, latent behaviors [78], [79].

By anchoring data provenance to the same cryptographic attestation framework used for model weights, integrity guarantees extend beyond file-level verification to encompass the semantic origins of model behavior. In practice, these guarantees can be further strengthened by committing cryptographic hashes of training datasets to an append-only transparency log prior to the commencement of training, preventing retroactive modification of provenance metadata and ensuring that data lineage is immutable from inception. These provenance guarantees establish trust in how a model was created, but must be preserved as the model is distributed and deployed across operational environments.

\subsection{Secure Model Distribution Under Attested Supply Chains}
While the prototype secures a single checkpoint at rest, real-world deployments involve model redistribution and downstream fine-tuning across multiple organizational and infrastructural boundaries. As models traverse registries, mirrors, or undergo downstream fine-tuning, attestation can be applied recursively, binding each derived model to the cryptographically verified state of its predecessor. This preserves a transitive chain of trust across distribution boundaries, ensuring that security guarantees remain intact as models evolve across operational environments. In such environments, traditional file-level checksums provide insufficient protection, as adversaries may substitute or subtly modify weights while preserving apparent functionality.

Extending the prototype into a full weight-attestation framework requires enforcing signature verification at each distribution boundary and rejecting unsigned or policy-noncompliant artifacts during deployment. This approach ensures that every executable model instance remains cryptographically bound to its originating training pipeline and approved configuration.

This mechanism contrasts with model watermarking, which embeds detectable patterns into model outputs for post-hoc attribution. While watermarking can assist in tracing misuse after deployment, it does not prevent tampered or malicious models from executing and can often be removed or weakened through fine-tuning [77]. Weight attestation, by contrast, enforces integrity prior to execution, aligning more directly with zero-trust supply-chain principles and providing enforceable pre-execution guarantees against checkpoint substitution and supply-chain tampering.

\subsection{Robustness Against Agentic LLM Attacks}
While weight attestation secures the static integrity of LLM artifacts, agentic LLM systems introduce a dynamic supply-chain surface at runtime. Tool access, persistent memory, and autonomous planning effectively extend the supply chain beyond model weights into execution-time dependencies [78].

From a systems perspective, agentic robustness represents the runtime continuation of build-time trust guarantees. Practical extensions of the attestation architecture include multi-stage plan validation, where an agent’s proposed action sequence is checked against dependency trust policies and privilege boundaries prior to execution. Similarly, tool invocation and external API access can be gated by signed manifests and runtime verification, preventing compromised tools from corrupting agent behavior.

Persistent memory introduces an additional integrity risk: malicious prompts or artifacts may poison agent state long after initial deployment. Addressing this risk requires constraint-aware memory management and continuous validation of stored context against policy and provenance metadata. Together, these mechanisms extend supply-chain assurance from static artifacts to autonomous behavior.

\subsection{From Proof-of-Concept to Scalable, Enforceable Assurance}
The presented prototype demonstrates that cryptographic weight attestation and verification are already achievable using existing tooling. Scaling this approach to frontier-scale models and continuously evolving agentic systems introduces open challenges, including transparency-log scalability, semantic binding of safety properties to cryptographic attestations, and continuous enforcement under model updates. For frontier-scale models, naive hashing of multi-gigabyte checkpoints can introduce significant I/O overhead that negatively impacts deployment latency. Merkle-tree–based partial attestation provides a practical refinement, enabling parallelized verification of model shards or layers without rehashing the entire artifact. This approach preserves integrity guarantees while mitigating verification costs, directly addressing scalability constraints in high-throughput deployment pipelines.

Crucially, these challenges build directly upon the PoC architecture described above. Rather than representing a speculative research agenda, they define a sequence of incremental engineering extensions that can be deployed, evaluated, and strengthened as LLM-based security systems scale. In this sense, LLM supply-chain security transitions from an abstract governance concern to a concrete systems problem with enforceable guarantees.

\subsection{Reframing Prior Cybersecurity Trajectories}
The proposed research agenda does not emerge in isolation, but rather builds upon and fundamentally reinterprets several foundational cybersecurity research trajectories. However, the integration of large language models into security-critical workflows violates key assumptions underlying prior approaches, necessitating new theoretical and methodological directions.

\begin{enumerate}
    \item \textbf{From Adversarial ML to Cyber Threat Inflation:} Traditional adversarial machine learning has largely focused on perturbation-based attacks and evasion of fixed classifiers under bounded threat models. These frameworks assume that adversarial effort scales linearly with attack complexity and that model misuse requires significant expertise. LLMs invalidate these assumptions by enabling the low-cost, automated generation of polymorphic malware, exploits, and social-engineering artifacts. As a result, the dominant challenge is no longer isolated evasion, but cyber threat inflation, where the marginal cost of producing diverse and adaptive attacks approaches zero. Addressing this shift requires research agendas that emphasize systemic resilience, rate-limiting of attack generation, and defenses robust to continuously evolving threat distributions rather than static adversarial examples.
    
    \item \textbf{From Zero-Trust to Semantic Trust:} Traditional Zero-Trust Architectures (ZTA) have traditionally centered on identity verification, authentication, and access control, operating under the assumption that software artifacts are static, human-authored, and auditable prior to execution. LLM-generated code and agentic behaviors violate these assumptions by introducing probabilistic, opaque, and dynamically synthesized artifacts whose intent may not be inferable from syntax or origin alone. In LLM-integrated systems, trust decisions must therefore extend beyond identity and provenance to encompass semantic trust: continuous verification of the intent, side effects, and policy compliance of AI-generated actions at run time. This shift motivates new research into semantic policy enforcement, intent verification, and dynamic trust evaluation for AI-driven systems.
    
    \item \textbf{From Static Robustness to Adaptive Resilience:} Conventional cyber defenses rely heavily on static signatures, fixed rule sets, and periodic retraining cycles, reflecting an assumption that threat evolution is incremental and observable. LLM-enabled attackers undermine this model by rapidly generating novel attack variants and adapting behaviors in response to deployed defenses. Similarly, LLM-based defensive systems may themselves evolve through continual learning, fine-tuning, or agentic feedback loops. These dynamics require a move toward adaptive and lifelong robustness, where defensive mechanisms continuously update their detection logic, threat models, and trust assumptions in response to both environmental changes and emergent supply-chain vulnerabilities.
\end{enumerate}

\section{Conclusion and Research Implications}
The integration of large language models into cybersecurity represents a structural shift in both the threat landscape and the defensive toolkit. As this paper has argued, LLMs simultaneously amplify defensive capabilities such as automated vulnerability discovery, code analysis, and security orchestration, while dramatically lowering the cost, expertise, and scale required for sophisticated cyberattacks. This dual-use dynamic accelerates cyber threat inflation, widens the asymmetry between attackers and defenders, and exposes new classes of systemic risk that cannot be addressed through incremental extensions of existing security frameworks.

This work makes two primary contributions that together motivate a focused research agenda. First, we introduce the LLM Scalability Risk Index (LSRI), which provides a structured lens for stress-testing the operational, economic, and compliance trade-offs associated with deploying LLMs in security-critical environments. Second, we develop a model-supply-chain security architecture that establishes a verifiable root of trust across data acquisition, training, and deployment, offering a unifying perspective on how vulnerabilities introduced at early stages can propagate and amplify downstream.

More broadly, this paper situates LLM-enabled cybersecurity challenges within the lineage of prior research agendas, including adversarial machine learning, zero-trust architectures, and cyber resilience. While these paradigms remain essential, LLMs fundamentally transform their underlying assumptions by introducing probabilistic generation, opaque decision-making, and autonomous action at scale. As a result, securing AI-integrated systems demands research that moves beyond isolated attack classes or defensive techniques toward a systems-oriented understanding of trust, robustness, and resilience across the entire LLM lifecycle. In this sense, the paper contributes towards shifting the community from descriptive taxonomies of LLM risks toward a coherent and actionable research program for AI-enabled cyber defense.

Additionally, the feasibility of AI governance is not purely a policy question, but is fundamentally constrained by the technical properties of LLM systems.

\subsection{Design Constraints for Feasible AI Governance}
Although governance mechanisms for LLM-enabled cybersecurity systems are often framed as regulatory or institutional challenges, the feasibility of governance for LLM-enabled security depends on system-level guarantees. Effectiveness requires alignment with verifiable, scalable technical mechanisms. Any practically feasible approach must satisfy three core constraints:

\begin{enumerate}
    \item \textbf{Verifiability:} Governance must rely on verifiable evidence rather than self-attestation. This necessitates auditing the LLM lifecycle, including provenance and fine-tuning through cryptographically verifiable artifacts like dataset commitments and weight attestation. Without these, oversight remains manual and unscalable.
    
    \item \textbf{Scalability:} Frameworks must match the scale of model size and autonomy. Ad hoc risk assessments cannot keep pace with continuous fine-tuning and reuse. Feasible governance requires automation-friendly mechanisms, such as secure aggregation and machine-verifiable compliance signals, to minimize human-intensive overhead.
    
    \item \textbf{Adaptivity:} Static compliance checklists cannot keep pace with agentic threat evolution. Effective governance must support Adaptive Enforcement, where trust assumptions---quantified via parametric models like the LSRI (Eq. \ref{eq:lsri}), evolve in response to real-time telemetry such as latency shifts or supply-chain integrity alerts. This ensures that governance is a dynamic runtime process rather than an offline audit.
\end{enumerate}

Taken together, these constraints highlight that governance feasibility is inseparable from advances in technical assurance. Rather than viewing governance as an external layer imposed on AI systems, these constraints emphasize the need for co-design between governance mechanisms and supply-chain security primitives, including provenance tracking, cryptographic integrity checks, and runtime monitoring.

This research agenda also highlights the need to formalize a standardized AI Model Bill of Materials (AI-BOM) to provide a machine-readable format for the signed metadata and provenance records proposed in this work, facilitating interoperability across security tooling and governance frameworks. In addition, future research should characterize the performance–security Pareto frontier, quantifying the trade-offs between cryptographic verification overhead and gains in integrity assurance. Understanding these trade-offs is essential for the practical adoption of enforceable supply-chain guarantees in large-scale AI deployments.

Beyond build-time and pre-execution controls, future research may explore complementary runtime assurance techniques for long-lived agentic sessions. One possible direction is the use of periodic heartbeat checks, where a secondary, lightweight verifier model audits an agent’s recent action plans against its signed safety policies. Importantly, such mechanisms are not substitutes for enforceable supply-chain guarantees, but potential secondary defenses for detecting semantic drift or policy violations that emerge during extended autonomous operation.

\section{Policy and Practice Implications}
While the primary contribution of this paper is research-facing, the proposed agenda carries important implications for practitioners, organizations, and policymakers. As large language models are increasingly embedded into security-critical workflows, governance and assurance mechanisms must evolve to account for risks arising from opaque training pipelines, third-party model dependencies, and autonomous agent behavior.

\begin{enumerate}
    \item \textbf{For Policymakers and Regulators:} Efforts should focus on establishing agile and internationally harmonized regulatory frameworks that encourage responsible AI innovation while mandating baseline security, transparency, and accountability standards for high-risk LLM applications in cybersecurity. Public–private partnerships play a critical role in enabling information sharing and aligning regulatory expectations with operational realities. For example, the UK National Cyber Security Centre’s recent enterprise-level guidance on LLM usage emphasizes prompt sanitization, API handling, and audit logging as practical risk-reduction measures [70].
    
    \item \textbf{For Security Organizations and Practitioners (CISOs, SecOps Teams):} Organizations should prioritize comprehensive strategies for integrating LLMs into security workflows, including rigorous testing and validation of AI-enabled tools, continuous red-teaming against AI-augmented threats, and workforce upskilling to manage model behavior, dependencies, and failure modes. Supply-chain visibility and runtime monitoring should be treated as core security functions rather than optional enhancements.
    
    \item \textbf{For LLM Developers and AI Researchers:} Developers should emphasize security-by-design principles throughout the LLM lifecycle, from data curation and training to deployment and monitoring. This includes investment in safer model architectures, bias detection and mitigation techniques tailored to security contexts, and robust mechanisms for content authenticity and provenance to counter AI-generated disinformation and malware.
\end{enumerate}

Ultimately, securing the future of LLM-enabled cybersecurity systems is not solely a technical challenge, but a socio-technical one that requires alignment between research, engineering practice, and governance. Key near-term actions for major stakeholder groups are summarized in Table \ref{tab:stakeholder_actions}.

\begin{table}[htbp]
\centering
\caption{Recommended immediate actions for key stakeholder groups to mitigate dual-use risks of LLMs.}
\label{tab:stakeholder_actions}
\vspace{2mm}
\begin{tabular}{lp{11cm}}
\toprule
\textbf{Stakeholder} & \textbf{What They Must Do Now} \\
\midrule
Governments & Enact dual-use-specific AI policies \& register frontier models \\
\addlinespace
Enterprises (CISOs) & Integrate LLMs into SecOps with real-time XAI + red teaming \\
\addlinespace
Researchers & Build verifiable benchmarks for LLM explainability and zero-day detection \\
\addlinespace
App Platforms & Embed LLMs into app review pipelines with FL + differential privacy \\
\addlinespace
LLM Developers & Secure training pipelines, sign checkpoints, and release model cards \\
\bottomrule
\end{tabular}
\end{table}

\end{document}